\documentclass{webofc}

\usepackage[varg]{txfonts}   
\usepackage{hyperref}
\usepackage{url}
\usepackage{cleveref}
\usepackage{xcolor}
\hypersetup{colorlinks=true,citecolor=blue,urlcolor=blue,linkcolor=blue}

\newcommand{\cp}{\ensuremath{{\cal CP}}}

\newcommand{\MS}{{\ensuremath{\overline{\text{MS}}}}\xspace}

\newcommand{\nn}{\nonumber}

\newcommand{\delone}{\delta^{(1)}}
\newcommand{\deltwo}{\delta^{(2)}}

\begin{document}
\begin{flushright}
\texttt{DESY-26-048}
\end{flushright}

\setlength{\footskip}{20pt}
\title{Refining two-loop corrections to trilinear Higgs couplings in the Two-Higgs-Doublet Model}
%
%

\author{\firstname{Johannes} \lastname{Braathen}\inst{1}\fnsep\thanks{\email{johannes.braathen@desy.de}, \textit{Speaker}} \and
        \firstname{Felix} \lastname{Egle}\inst{1}\fnsep\thanks{\email{felix.egle@desy.de}} \and
        \firstname{Alain} \lastname{Verduras Schaeidt}\inst{1}\fnsep\thanks{\email{alain.verduras@desy.de}}
}

\institute{Deutsches Elektronen-Synchrotron DESY, Notkestr.~85,  22607  Hamburg,  Germany
          }

\abstract{
The precise determination of the Higgs self-couplings is an essential task for understanding electroweak symmetry breaking and probing physics beyond the Standard Model (SM). The calculation of two-loop corrections to scalar couplings is important as it provides a critical test of the perturbative stability of the theoretical predictions, especially in scenarios with extended scalar sectors where large one-loop corrections can occur. Moreover, two-loop corrections need to be taken into account for the future perspective of precisely measuring the trilinear Higgs self-coupling. We present new results for the leading two-loop corrections to trilinear Higgs couplings in the Two-Higgs-Doublet Model (2HDM). We focus in particular on the couplings $\lambda_{hhh}$ and $\lambda_{hhH}$, which are relevant for Higgs pair production at the (HL-)LHC or at future linear colliders. We address the renormalisation of the alignment limit in the Higgs basis and give some insights into technical details of the calculation. Finally, we discuss the phenomenological impact of our results on di-Higgs production differential distributions.
}
\maketitle
\pagestyle{plain}

\vfill\noindent\textit{Talk presented at the International Workshop on Future Linear Colliders 2025 (LCWS2025), Valencia, Spain}

\newpage

\section{Introduction}
\label{intro}
The trilinear Higgs self-coupling, $\lambda_{hhh}$, is a crucial probe of the shape of the Higgs potential realised in Nature and of the dynamics of the electroweak phase transition (EWPT). It is also a prime target to search for phenomena beyond the Standard Model (BSM). Precise measurements of $\lambda_{hhh}$, or rather its coupling modifier $\kappa_\lambda\equiv \lambda_{hhh}/\lambda_{hhh}^{\text{SM},\ (0)}$ (where $\lambda_{hhh}^{\text{SM},\ (0)}$ is the tree-level SM prediction), are thus flagship measurements at the LHC, its high-luminosity upgrade (HL-LHC) and future colliders. Currently the best combination from ATLAS and CMS~\cite{CMS:2026nuu} yields the bounds $-0.71<\kappa_\lambda<6.1$ at 95\% C.L., and significant improvements are expected at future colliders (see e.g.\ Refs.~\cite{ATLAS:2025eii,LinearColliderVision:2025hlt}). 

Sizeable deviations in $\lambda_{hhh}$ from its Standard Model (SM) value are known to occur commonly in many BSM theories with extended scalar sectors~\cite{Kanemura:2002vm,Kanemura:2004mg,Aoki:2012jj,Kanemura:2015fra,Kanemura:2015mxa,Arhrib:2015hoa,Kanemura:2016sos,Kanemura:2016lkz,He:2016sqr,Kanemura:2017wtm,Kanemura:2017gbi,Chiang:2018xpl,Basler:2018cwe,Senaha:2018xek,Braathen:2019pxr,Braathen:2019zoh,Kanemura:2019slf,Braathen:2020vwo,Basler:2020nrq,Bahl:2022jnx,Bahl:2022gqg,Falaki:2023tyd,Bahl:2023eau,Aiko:2023nqj,Cherchiglia:2024abx,Basler:2024aaf,Bahl:2025wzj,Braathen:2025qxf,Braathen:2025svl}, due to radiative corrections involving the BSM scalars. 
In Refs.~\cite{Senaha:2018xek,Braathen:2019pxr,Braathen:2019zoh}, two-loop corrections to $\lambda_{hhh}$ were computed for the first time in several BSM models, including the Two-Higgs-Doublet Model (2HDM), thereby confirming the perturbative convergence of the large corrections that can occur from the one-loop order onwards (see also Refs.~\cite{Brucherseifer:2013qva,Muhlleitner:2015dua,Braathen:2020vwo,Borschensky:2022pfc,Bahl:2025wzj,Dao:2025wqs} for other two-loop computations). Moreover, it was shown in Ref.~\cite{Bahl:2022jnx} that the comparison of high-precision predictions for $\lambda_{hhh}$ with the experimental limits derived from di-Higgs production searches can provide a powerful tool to probe or constrain the parameter space of BSM theories with extended scalar sectors, beyond the reach of other state-of-the-art experimental and theoretical constraints. In this context, the inclusion of the known and numerically significant two-loop corrections to $\lambda_{hhh}$ is essential for reliable interpretations of the experimental bounds in terms of the allowed or excluded regions of BSM parameter space.  

The trilinear self-coupling of the detected Higgs boson is however not the only trilinear scalar coupling required to reconstruct the shape of the Higgs potential in models with extended scalar sectors. Other trilinear couplings can appear in pair production processes of the detected Higgs boson, like $gg\to hh$ at the (HL-)LHC, and can also receive significant radiative corrections, see e.g.\ Refs.~\cite{Arco:2025pgx,Braathen:2025qxf}. In the 2HDM, this is in particular the case of the coupling $\lambda_{hhH}$. 

Recent improvements~\cite{Degrassi:2023eii,Bahl:2025wzj,Degrassi:2025pqt} in the calculation of trilinear scalar couplings at two loops and in the renormalisation of the condition of alignment now motivate revisiting the earlier results of Refs.~\cite{Braathen:2019pxr,Braathen:2019zoh} for $\lambda_{hhh}$ in order to relax some of the simplifications made therein (in particular neglecting loop-induced corrections to the alignment condition on the \cp-even mixing angle $\alpha$), and to extend the scope of these calculations to other couplings such as $\lambda_{hhH}$. 
In these proceedings, we present preliminary results from our ongoing work on improving predictions for the leading two-loop corrections to the trilinear scalar couplings contributing to Higgs pair production in the 2HDM.

\section{The Two-Higgs-Doublet Model}
We present in this section the setting of our work in the \cp-conserving 2HDM. We start by introducing the scalar potential
\begin{align}
V(\Phi_{\mathrm{SM}},\Phi_{\mathrm{BSM}})=&\ M_{11}^2|\Phi_{\mathrm{SM}}|^2 + M_{22}^2|\Phi_{\mathrm{BSM}}|^2 - M_{12}^2(\Phi_{\mathrm{BSM}}^{\dagger}\Phi_{\mathrm{SM}}+\mathrm{ h.c.}) \nn\\
&+ \frac{\Lambda_1}{2}|\Phi_{\mathrm{SM}}|^4+\frac{\Lambda_2}{2}|\Phi_{\mathrm{BSM}}|^4 +\Lambda_3|\Phi_{\mathrm{SM}}|^2|\Phi_{\mathrm{BSM}}|^2+\Lambda_4|\Phi_{\mathrm{BSM}}^{\dagger}\Phi_{\mathrm{SM}}|^2 \nn \\ 
&+ \left[\frac{\Lambda_5}{2}(\Phi_{\mathrm{BSM}}^{\dagger}\Phi_{\mathrm{SM}})^2+(\Lambda_6|\Phi_{\mathrm{SM}}|^2+\Lambda_7|\Phi_{\mathrm{BSM}}|^2)\Phi_{\mathrm{SM}}^{\dagger}\Phi_{\mathrm{BSM}}+\text{ h.c.}\right]\,,
\end{align}
written in terms of the two doublets $\Phi_{\mathrm{SM}}$ and $\Phi_{\mathrm{BSM}}$ defined as
\begin{align}
    \Phi_{\mathrm{SM}}&=\left(
\begin{matrix}
G'^+\\
\frac{1}{\sqrt{2}}(v+\phi_{\mathrm{SM}}+iG'^0)
\end{matrix}\right),\quad\text{  and  }\quad\Phi_{\mathrm{BSM}}=\left(
\begin{matrix}
H'^+\\
\frac{1}{\sqrt{2}}( v_{\mathrm{BSM}} + \phi_{\mathrm{BSM}}+iA')
\end{matrix}\right) \,,
\end{align}
and with all masses and quartic couplings taken to be real. 
This is the most general form of the potential for a \cp-conserving 2HDM, but as the notation indicates we intend to work in the Higgs basis, where the doublet $\Phi_{\mathrm{SM}}$ contains the SM-like scalars and the electroweak (EW) vacuum expectation value (VEV) $v$. This implies that $v_\mathrm{BSM}=0$ at tree level; although we note that this relation has to be enforced at loop level as well. In this basis the (Higgs) alignment limit is simply obtained by taking $\Lambda_6 \to 0$. 

Moreover, the Yukawa sector couplings to the top quark are given by
\begin{align}
\mathcal{L}_{\mathrm{Yuk}}&\supset - \bar{Q}_\mathrm{L} Y_{\mathrm{SM}}^{t} \tilde{\Phi}_{\mathrm{SM}} t_{\mathrm{R}} - \bar{Q}_\mathrm{L} Y_{\mathrm{BSM}}^{t} \tilde{\Phi}_{\mathrm{BSM}} t_{\mathrm{R}}\,, \nn\\
\text{where } &Q_\mathrm{L}= \begin{pmatrix}
 t_\mathrm{L} & b_\mathrm{L} \end{pmatrix}^\mathrm{T}\,,\quad \text{and}\quad Y_{\mathrm{BSM}}^{t}=\zeta_t Y_{\mathrm{SM}}^{t}\,.
\end{align}
where we introduced an a-priori generic coupling modifier $\zeta_t$, as in a flavour-aligned case. 

Next, we introduce the tadpole parameters, corresponding to the two minimisation conditions
\begin{subequations}
\begin{align}
\left. \frac{\partial V_{\mathrm{2HDM}}}{\partial \phi_{\mathrm{SM}}} \right|_{\{ \phi \} =0} &\stackrel{!}{=} T_{\phi_{\mathrm{SM}}} \quad (\phi = \phi_{\mathrm{SM}}, \phi_{\mathrm{BSM}},G'^0,A'^0,G'^\pm,H'^\pm)\,, \\
\left. \frac{\partial V_{\mathrm{2HDM}}}{\partial \phi_{\mathrm{BSM}}} \right|_{\{ \phi \} =0} &\stackrel{!}{=} T_{\phi_{\mathrm{BSM}}} \quad (\phi = \phi_{\mathrm{SM}}, \phi_{\mathrm{BSM}},G'^0,A'^0,G'^\pm,H'^\pm)\,,
\end{align}
\end{subequations}
which have to vanish in order for our chosen VEV configuration to be a minimum of the potential. The vanishing of the tadpoles also has to be enforced at loop level.

In the standard approach, one would introduce at this point mixing angles --- $\alpha^\prime$ and $\beta^\prime$ --- in order to define the mass eigenstates via the rotations
\begin{align}
    \begin{pmatrix}
H \\ h
\end{pmatrix}&= R_{\alpha'}^{\mathrm{T}} \begin{pmatrix}
\phi_{\mathrm{SM}} \\ \phi_{\mathrm{BSM}}\end{pmatrix} \,, \, \begin{pmatrix}
G^0 \\ A
\end{pmatrix}= R_{\beta'}^\mathrm{T} \begin{pmatrix}
G'^0 \\ A'
\end{pmatrix} \,, \, \begin{pmatrix}
G^+ \\ H^+
\end{pmatrix}= R_{\beta'}^\mathrm{T} \begin{pmatrix}
G'^+ \\ H'^+
\end{pmatrix}\,,
\end{align}
where 
\begin{align}
    R_{\theta}\equiv\begin{pmatrix}\cos{\theta} & -\sin{\theta}\\ \sin{\theta} & \cos{\theta} \end{pmatrix}\,.
\end{align}
The mixing angle for the \cp-odd and charged sectors can be expressed with the usual relation 
\begin{align}
    \tan\beta^\prime=\frac{v_\text{BSM}}{v}\underset{\text{Higgs basis}}{=}0\,,
\end{align}
while the \cp-even mixing angle $\alpha^\prime$ can be related to the components of the $2\times 2$ \cp-even mass matrix according to
\begin{align}
\label{eq:alpharelation}
\mathcal{M}_{\phi_{\mathrm{SM}}\phi_{\mathrm{BSM}}}&=\begin{pmatrix}
\mathcal{M}_{11} & \mathcal{M}_{12} \\
\mathcal{M}_{12} & \mathcal{M}_{22}
\end{pmatrix} \quad \Rightarrow \quad\tan 2 \alpha' = \frac{2 \mathcal{M}_{12}}{\mathcal{M}_{11}- \mathcal{M}_{22}}\,.
\end{align}
As we will use below, the tadpole parameters can also be transformed under this base rotation, as 
\begin{align}
    \begin{pmatrix}
    T_{\phi_{\mathrm{SM}}} \\ T_{\phi_{\mathrm{BSM}}}
    \end{pmatrix} &= R_{\alpha'} \begin{pmatrix}
    T_H \\ T_h
    \end{pmatrix}\,, 
\end{align}
$T_h$ and $T_H$ being the tadpole parameters in the mass basis. 

However, instead of this standard approach, we follow here the approach of Ref.~\cite{Degrassi:2023eii}, and omit the introduction of mixing angles (we however keep a finite rotation of $\alpha'=-\frac{\pi}{2}$ to maintain the same convention as is commonly used for the 2HDM). This can be done since we work in the alignment limit and thus the mass matrices are already diagonal at tree level. In turn, this choice implies that we have to define the counterterms for $\Lambda_6$ and $v_\mathrm{BSM}$ by demanding that the off-diagonal contributions to the scalar mass matrices are cancelled. While this approach is entirely equivalent to the one with mixing angles, this alternative approach has the advantage that it does not require the introduction of mixing angles that are afterwards set to fixed values ($0$ or $-\pi/2$). At the same time, all UV divergences can still be eliminated, thus making this approach simpler and more convenient for our calculations. 

Finally, to make contact with the $\mathbb{Z}_2$-symmetric version of the 2HDM, our potential parameters ($M_{ij}^2$ and $\Lambda_i$) can be expressed in terms of input parameters in the $\mathbb{Z}_2$ basis (see e.g. Refs.~\cite{Branco:2011iw,Bernon:2015qea,Bernon:2015wef}). As a consequence, the Higgs-basis quartic couplings have to obey the relations
\begin{subequations}
\label{eq:HBzoZ2}
\begin{align}
\Lambda_2 &= \Lambda_1 +2 (\Lambda_6+ \Lambda_ 7) \cot 2 \beta\,, \\
\Lambda_3+\Lambda_4+\Lambda_5&= \Lambda_1+2 \Lambda_6 \cot2 \beta - \frac{(\Lambda_6-\Lambda_7)}{\cot2 \beta}\,,
\end{align}
\end{subequations}
where $\beta$ denotes the mixing angle in the \cp-odd and charged sectors in the $\mathbb{Z}_2$ basis. Finally, the coupling modifier $\zeta_t$ can in the $\mathbb{Z}_2$-symmetric case be expressed as
\begin{align}
    \zeta_t=\cot \beta\,.
\end{align}

\section{Calculational setup and renormalisation scheme}
We describe now the setup of our calculation in the aligned 2HDM, with a particular emphasis on the renormalisation procedure that is carried out. 

The aim of the work presented here is to determine the magnitude of the dominant two-loop BSM corrections to the trilinear scalar couplings relevant for Higgs pair production. Therefore, we compute purely scalar as well as mixed scalar-fermion two-loop contributions to the $hhh$ and $hhH$ three-point functions, in the limit of vanishing external momenta (discussions of the relevance of external-momentum effects in BSM scenarios like the 2HDM can be found e.g.\ in Ref.~\cite{Bahl:2023eau}). Additionally, we neglect subleading contributions from light scalars, and for this purpose we set $m_h\to 0$ and work in the gaugeless limit.  

The calculation of the two-loop corrections is performed in parallel with two different approaches:
\begin{itemize}
\item In a first, diagrammatic, approach, relevant genuine two-loop and subloop renormalisation diagrams are generated with \texttt{FeynArts}~\cite{Hahn:2000kx} using a model file created for the Higgs basis of the 2HDM with \texttt{SARAH}~\cite{Staub:2009bi,Staub:2010jh,Staub:2012pb,Staub:2013tta}. The corresponding amplitudes are then computed and simplified using \texttt{FeynCalc}~\cite{MERTIG1991345,Shtabovenko:2016sxi,Shtabovenko:2020gxv,Shtabovenko:2023idz}, and lastly the reduction of the involved loop integrals to a set of master integrals is done with \texttt{Tarcer}~\cite{Mertig:1998vk}. 
\item In a second approach, we compute the effective potential up to the leading two-loop level in terms of contributions involving BSM scalars and the top quark, applying generic results (expressed in terms of \MS renormalised parameters) from the literature~\cite{Martin:2001vx} to the 2HDM in the Higgs basis. We then take field derivatives with respect to the $h$ and $H$ fields to obtain expressions for the two-loop corrections to the trilinear couplings. These are then also expressed in terms of \MS-renormalised parameters, but a scheme conversion can be performed using the finite parts of the counterterms discussed below. 
\end{itemize}
One of the strongest cross-checks of our work is that we have verified that we can obtain the same results using both approaches. 

Turning next to the renormalisation of the 2HDM, there are ten parameters that require counterterms for our computations, namely (in the Higgs basis)
\begin{align}
    T_h,\, T_H,\, v,\, v_\text{BSM},\, m_h,\, m_H,\, m_A,\, m_{H^\pm},\, M_{22},\, \Lambda_6,\, \Lambda_7\,.
\end{align}
We recall that $T_h=T_H=0$ due to the minimisation conditions of the potential, $m_h\to0$ from our approximation, $\Lambda_6=0$ as we consider the alignment limit of the 2HDM, and $v_\text{BSM}=0$ as we are in the Higgs basis. Nevertheless, these five vanishing parameters still require renormalisation. 
Given the approximations stated above, two-loop counterterms are only needed for $T_h$, $T_H$, $m_h$, $v_\text{BSM}$, and $\Lambda_6$, while for the other parameters one-loop counterterms are sufficient (because these parameters only enter expressions of $\lambda_{hhh}$ and $\lambda_{hhH}$ from the one-loop level). Furthermore, because $m_h=0$ and $\Lambda_6=0$, the tree-level predictions for both $\lambda_{hhh}$ and $\lambda_{hhH}$ vanish, which in turn means that we only need one-loop field renormalisation constants. We note that the classes of dominant one- and two-loop corrections to $\lambda_{hhh}$ and $\lambda_{hhH}$ computed in this work involve different couplings than the tree-level predictions, and are thus independent from them. 

In the following, we discuss the renormalisation conditions used for each of these parameters and field-renormalisation constants. To clarify the notations, we use throughout this section $\delone_\text{CT} p$ and $\deltwo_\text{CT} p$ in order to denote respectively the one- and two-loop counterterms of a given parameter $p$.
\medskip

\noindent\textbf{Tadpole parameters:} the tadpole parameters $T_h$ and $T_H$ are renormalised using the standard ``parameter-renormalised tadpole scheme'' (PRTS, see e.g.\ Ref.~\cite{Denner:1991kt}), in which we require that the renormalised tadpoles vanish order-by-order in perturbation theory.
\medskip

\noindent\textbf{Masses and diagonal wave-function renormalisation:} 
The renormalisation of the scalar masses --- $m_h$, $m_H$, $m_A$, $m_{H^\pm}$ --- is also done using the usual on-shell (OS) scheme. We do not reproduce the entire relations for the mass counterterms here for brevity, but we  refer instead the reader to e.g.\ Refs.~\cite{Denner:1991kt,Krause:2016oke}
for more details. 

Moving to the field renormalisation of the two \cp-even Higgs fields, $h$ and $H$, we apply the renormalisation transformation
\begin{align}
\begin{pmatrix}
H \\ 
h
\end{pmatrix} \to \sqrt{1+\delta_\text{CT} Z} \begin{pmatrix}
H \\ h
\end{pmatrix} \approx \left( 1 + \frac{\delta_\text{CT}Z}{2} - \frac{\delta_\text{CT} Z^2}{8} \right) \begin{pmatrix}
H \\ h
\end{pmatrix}\,,
\end{align}
where $\delta_\text{CT} Z$ is now a matrix, defined as
\begin{align}
\label{eq:dCT_Z}
\delta_\text{CT} Z =\begin{pmatrix}
\delta_\text{CT} Z_{HH} & \delta_\text{CT} Z_{Hh} \\
\delta_\text{CT} Z_{hH} & \delta_\text{CT} Z_{hh}
\end{pmatrix}\,.
\end{align}
The diagonal elements of $\delta_\text{CT} Z$ are also defined in the usual OS scheme. Similarly, the diagonal $\delta_\text{CT}Z_{AA}$ required below (c.f.\ the discussion of $\Lambda_7$) is also defined with a standard OS scheme.
\medskip

\noindent\textbf{Off-diagonal wave-function renormalisation and renormalisation of the alignment condition ($\Lambda_6$):} 
As mentioned in the previous section we define the counterterms for $\Lambda_6$ 
by demanding that it cancels the off-diagonal contributions to the \cp-even mass matrix at zero external momentum. 

Having set $v_\mathrm{BSM}=0$, this matrix reads at the tree level
\begin{subequations}
\begin{align}
D_{\{
\phi_{\mathrm{SM}}\phi_{\mathrm{BSM}}\}}&= \begin{pmatrix}
m_h^2 & \Lambda_6 v^2 \\
\Lambda_6 v^2 & m_H^2
\end{pmatrix} + \begin{pmatrix}
\frac{T_{\phi_{\mathrm{SM}}}}{v} & \frac{T_{\phi_{\mathrm{BSM}}}}{v} \\
\frac{T_{\phi_{\mathrm{BSM}}}}{v} & 0 
\end{pmatrix}\,,
\end{align}
\end{subequations}
and after applying the replacement $\Phi_{\mathrm{BSM}} \to -H$ and $T_{\phi_\text{BSM}} \to - T_H$ (i.e.\ rotating with an angle $\alpha^\prime=-\frac{\pi}{2}$), we have
\begin{align}
D_{\{Hh\}}&= \begin{pmatrix}
m_H^2 & -\Lambda_6 v^2 \\
-\Lambda_6 v^2 & m_h^2
\end{pmatrix} + \underbrace{\begin{pmatrix}
0 & -\frac{T_H}{v} \\
-\frac{T_H}{v} & \frac{T_h}{v} 
\end{pmatrix}}_{\equiv T_{\{Hh\}}}\,.
\end{align}
Demanding now that the off-diagonal elements of the loop-corrected \cp-even scalar mass matrix vanish, we obtain expressions that we can solve for $\delta_\text{CT}\Lambda_6$ and the off-diagonal $\delta_\text{CT} Z$. There is some degree of redundancy in this context, as the counterterm for $\Lambda_6$ shares the same role as $\delta_\text{CT} Z_{Hh}$ and we can thus set the latter to zero. We then obtain the relations
\begin{subequations}
\begin{align}
\Sigma_{Hh}^{(1)}(0)&= \delta^{(1)}_\text{CT} D_{Hh}\\
\Sigma_{Hh}^{(2)}(0)&= \delta^{(2)}_\text{CT} D_{Hh} +\frac{1}{2} \delta^{(1)}_\text{CT} D_{hh} \delta^{(1)}_\text{CT} Z_{hH} +\delta^{(1)}_\text{CT} D_{Hh}\left( \frac{\delta^{(1)}_\text{CT} Z_{hh} + \delta^{(1)}_\text{CT} Z_{HH}}{2} \right) \\
\Sigma_{Hh}^{(1)}(m_H^2)&=\delta^{(1)}_\text{CT} D_{Hh} - \frac{1}{2} m_H^2\delta^{(1)}_\text{CT} Z_{hH} 
\end{align}
\end{subequations}
Solving for the counterterm of $\Lambda_6$, at one- and two-loop orders, we obtain
\begin{subequations}
\begin{align}
\delta^{(1)}_\text{CT} \Lambda_6&=-\frac{1}{v^2} \left(\Sigma_{Hh}^{(1)}(0) - \delta^{(1)}_\text{CT} T_{Hh} \right)\,,\\
\delta^{(2)}_\text{CT} \Lambda_6&=-\frac{1}{v^2} \left[\Sigma_{Hh}^{(2)}(0) - \delta ^{(2)}_\text{CT}T_{Hh} - (-v^2 \delta^{(1)}_\text{CT} \Lambda_6 +\delta^{(1)}_\text{CT} T_{Hh}) \left(\frac{\delta^{(1)}_\text{CT} Z_{hh} + \delta^{(1)}_\text{CT} Z_{HH}}{2}\right) \right. \\
& \left. \qquad \qquad -\frac{1}{2}\delta^{(1)}_\text{CT} D_{hh}\delta^{(1)}_\text{CT} Z_{hH} +2v\,\delta^{(1)}_\text{CT} \Lambda_6 \delta^{(1)}_\text{CT} v \right]\,. \nonumber
\end{align}
\end{subequations}
Furthermore, we find that
\begin{align}
\delta_\mathrm{CT}^{(1)} Z_{hH}&=-\frac{2}{m_H^2} \left( \Sigma_{Hh}^{(1)}(m_H^2) + \delta^{(1)}_\text{CT} \Lambda_6 v^2 -\delta_\text{CT}^{(1)} T_{Hh} \right)=-\frac{2}{m_H^2} \left( \Sigma_{Hh}^{(1)}(m_H^2) - \Sigma_{Hh}^{(1)}(0) \right)\,, 
\end{align}
and similarly for the \cp-odd case (needed below for the renormalisation of $\Lambda_7)$, we have
\begin{align}
    \delta_\mathrm{CT}^{(1)} Z_{G^0A}&= -\frac{2}{m_A^2} \left( \Sigma_{G^0A}^{(1)}(m_A^2) - \Sigma_{G^0A}^{(1)}(0) \right)\,.
\end{align}
\medskip

\noindent\textbf{Renormalisation of the electroweak VEV $v$:}
The electroweak VEV $v$ can be renormalised as in the SM. The VEV counterterm then depends on the $W$- and $Z$-boson mass counterterms $\delta_\mathrm{CT}m_W^2$ and $\delta_\mathrm{CT}m_Z^2$ (for which we adopt an OS scheme) and on the counterterm for the electric charge $\delta_\mathrm{CT}Z_e$. We follow here Ref.~\cite{Bredenstein:2006rh} to define the electric charge in terms of the Fermi constant $G_\mu$. It should be noted that we only include terms that remain after taking the gaugeless limit and the limit $m_h \to 0$.
\medskip

\noindent\textbf{Renormalisation of $v_\text{BSM}$ (Higgs basis condition):} 
Similarly to our scheme for $\Lambda_6$, we define the counterterm for $v_\text{BSM}$ such that it cancels the off-diagonal contribution to the pseudoscalar (i.e.\ \cp-odd) mass matrix at zero external momentum. It, however, follows from the Goldstone theorem that the pseudoscalar (as well as the charged scalar) mass matrix are automatically diagonal for $p^2=0$, which means that we can set $\delta_\text{CT}v_\text{BSM}=0$ in our work. 
\medskip

\noindent\textbf{Renormalisation of $\zeta_t$:}
Comparing with the $\mathbb{Z}_2$-symmetric basis, we have the relation
\begin{align}
\zeta_t&= \cot(\beta)  \quad \Rightarrow \quad \zeta_t + \delta_{\mathrm{CT}} \zeta_t = \cot\left(\beta + \delta_{\mathrm{CT}} \beta\right)\,,
\end{align}
and therefore for the one-loop $\zeta_t$ counterterm we find
\begin{align}
\delta_{\mathrm{CT}}^{(1)} \zeta_t = - (1+\cot^2\beta)\, \delta_{\mathrm{CT}}^{(1)} \beta\,.
\end{align}
We can use the rigid symmetry approach (compare e.g.\ to Refs.~\cite{Denner:2016etu,Altenkamp:2017ldc}) to obtain a counterterm for $\beta$. In this approach, one only introduces diagonal WFR constants $Z_{\Phi_1}$, $Z_{\Phi_2}$ for the doublets in the $\mathbb{Z}_2$ basis and relates them to the WFR constants from the mass basis via a rotation. As we do not have any other mixing angles, this rotation is only given by the rotation from the mass basis to the Higgs basis, i.e. with the mixing angle $\beta$ (and a fixed rotation of $\alpha'=-\frac{\pi}{2}$) and we obtain the relation
\begin{align}
\sqrt{1+\delta_{\mathrm{CT}}Z} \begin{pmatrix}
H \\ h
\end{pmatrix}&= R_{\alpha' + \beta+\delta_{\mathrm{CT}} \beta}^\mathrm{T} \begin{pmatrix}
\sqrt{Z_{\Phi_1}} & 0 \\ 0 & \sqrt{Z_{\Phi_2}}
\end{pmatrix}R_{\alpha' + \beta}
\begin{pmatrix}
H \\ h
\end{pmatrix}\,,
\end{align}
where we now introduced a counterterm for $\beta$ and $\delta_{\mathrm{CT}}Z$ is the matrix defined in \cref{eq:dCT_Z}. We then find
\begin{align}
\delta_{\mathrm{CT}}^{(1)} \beta&=\frac{\delta_{\mathrm{CT}}^{(1)} Z_{Hh} - \delta_{\mathrm{CT}}^{(1)} Z_{hH}}{4}  \,,
\end{align}
from which we can in turn derive $\delta_{\mathrm{CT}}^{(1)}\zeta_t$. 
\medskip

\noindent\textbf{Renormalisation of the BSM mass parameter $M_{22}^2$:} Next, we define the counterterm $\delta_\text{CT} M_{22}^2$ by demanding that the decoupling of the two-loop BSM contributions to $\lambda_{hhh}$ be apparent when using a tree-level relation between $M_{22}^2$ and the BSM scalar masses 
(see also Refs.~\cite{Braathen:2019pxr,Braathen:2019zoh,Degrassi:2023eii}). In other words, we demand that 
\begin{align}
\delta^{(2)}\lambda_{hhh}|^\text{BSM}
\xrightarrow{M_{22}^2 \to \infty} 0 \,, 
\end{align}
when using relations of the form $m_{\Phi}^2 \simeq M_{22}^2 + \lambda_\Phi v^2 \,\, (\Phi=H,A,H^\pm)$ for the BSM scalars. From this condition, we obtain 
\begin{align}
    \delta^{(1)}_\text{CT} M_{22}^2 = \frac{3M_{22}^2}{16 \pi^2} \left[ \Lambda_2 \left(\frac{1}{\epsilon}+1 - \log\frac{M_{22}^2}{Q^2}\right) + 
    \frac{2\zeta_t^2m_t^2}{v^2}\left(\frac{1}{\epsilon}+2 - \log\frac{M_{22}^2}{Q^2}\right) \right]\,.
\end{align}
\medskip

\noindent\textbf{Renormalisation of the Lagrangian quartic coupling $\Lambda_7$:} 
Finally, to obtain the counterterm for $\Lambda_7$, we define an OS condition for the three-point function corresponding to the trilinear scalar coupling $\lambda_{HAA}$. Working at the one-loop order, we specifically demand that the renormalised one-loop value of this coupling should be equal to its tree-level value (similar schemes employing three-point functions for the OS renormalisation of extended scalar sectors have recently been devised in Refs.~\cite{Braathen:2025qxf,Bahl:2026anyhh}; see also Refs.~\cite{CoarasaPerez:1996xba,Freitas:2002um,Krause:2016oke,Denner:2018opp,Azevedo:2021ylf,Egle:2022wmq,Egle:2023pbm}). 
Written as an equation, this condition reads 
\begin{equation}
    \lambda_{HAA}^{(1)}\stackrel{!}{=}\lambda_{HAA}^{(0)}\,.
\end{equation}
Expanding the renormalised one-loop result for $\lambda_{HAA}^{(1)}$, we obtain the following equation, which can be used to extract the expression of $\delta^{(1)}_\text{CT} \Lambda_7$, 
\begin{align}
    &\delta^{(1)}\lambda_{HAA}^{\mathrm{vertex}} + \delta^{(1)}\lambda_{HAA}^{\mathrm{ext.-leg}} + \Lambda_7 \delta^{(1)}_\text{CT} v + \delta^{(1)}_\text{CT} \Lambda_7v\nn\\ 
    &+ \delta^{(1)}_\text{CT} Z_{G^0A} \frac{m_H^2-m_A^2}{v} + \delta^{(1)}_\text{CT} Z_{hH}\frac{M_{22}^2 - m_A^2}{v} + \Lambda_7 v  \left(\delta^{(1)}_\text{CT} Z_{AA} + \frac{1}{2}\delta^{(1)}_\text{CT} Z_{HH}\right)\stackrel{!}{=}0\,,
\end{align}
where $\delta^{(1)}\lambda_{HAA}^{\mathrm{vertex}}$ and $\delta^{(1)}\lambda_{HAA}^{\mathrm{ext.-leg}}$ are the one-loop vertex and external-leg corrections to $\lambda_{HAA}$ respectively.

\section{Leading two-loop results for $\lambda_{hhh}$ and $\lambda_{hhH}$}

\begin{figure}[ht]
    \centering
    \includegraphics[width=0.49\textwidth]{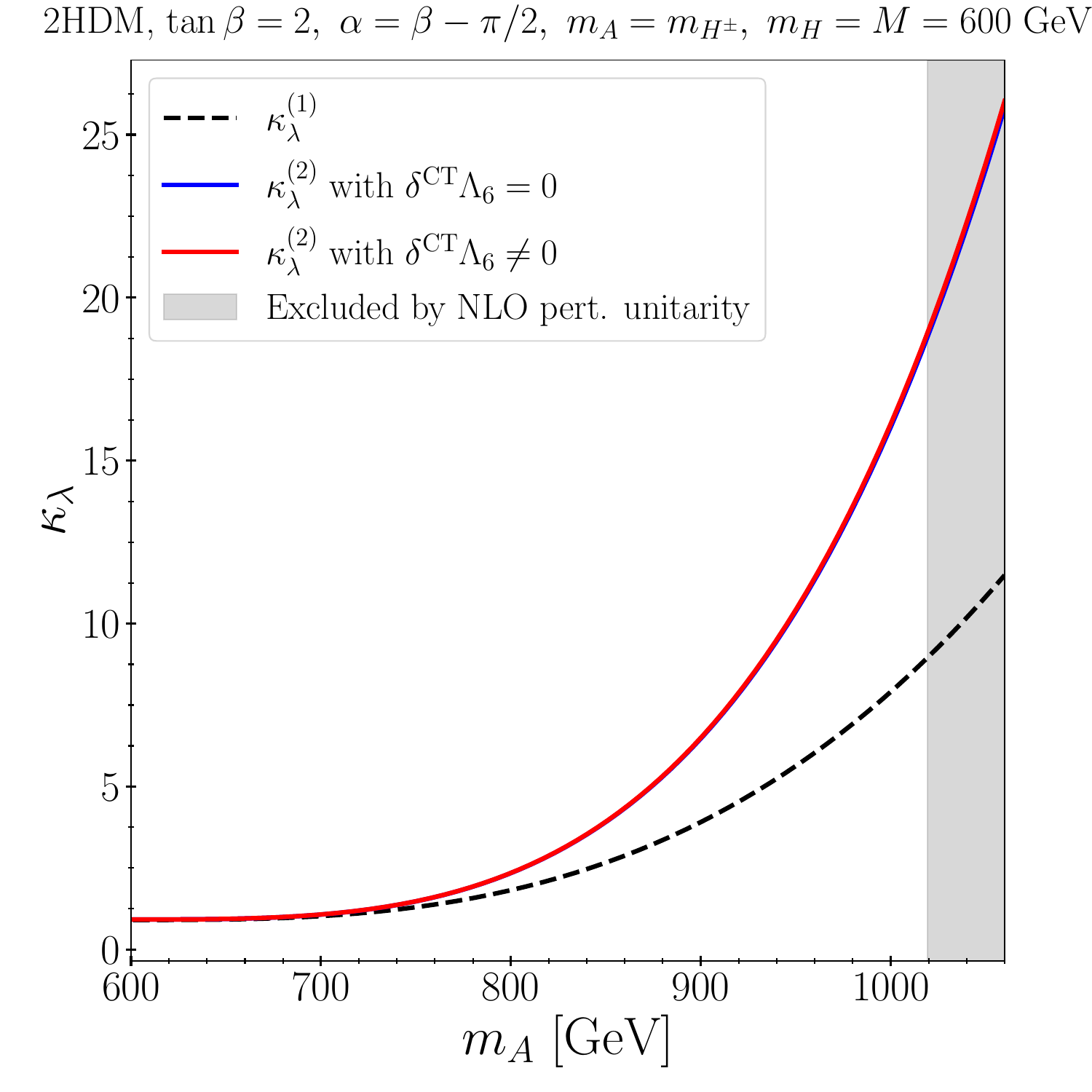}
    \includegraphics[width=0.49\textwidth]{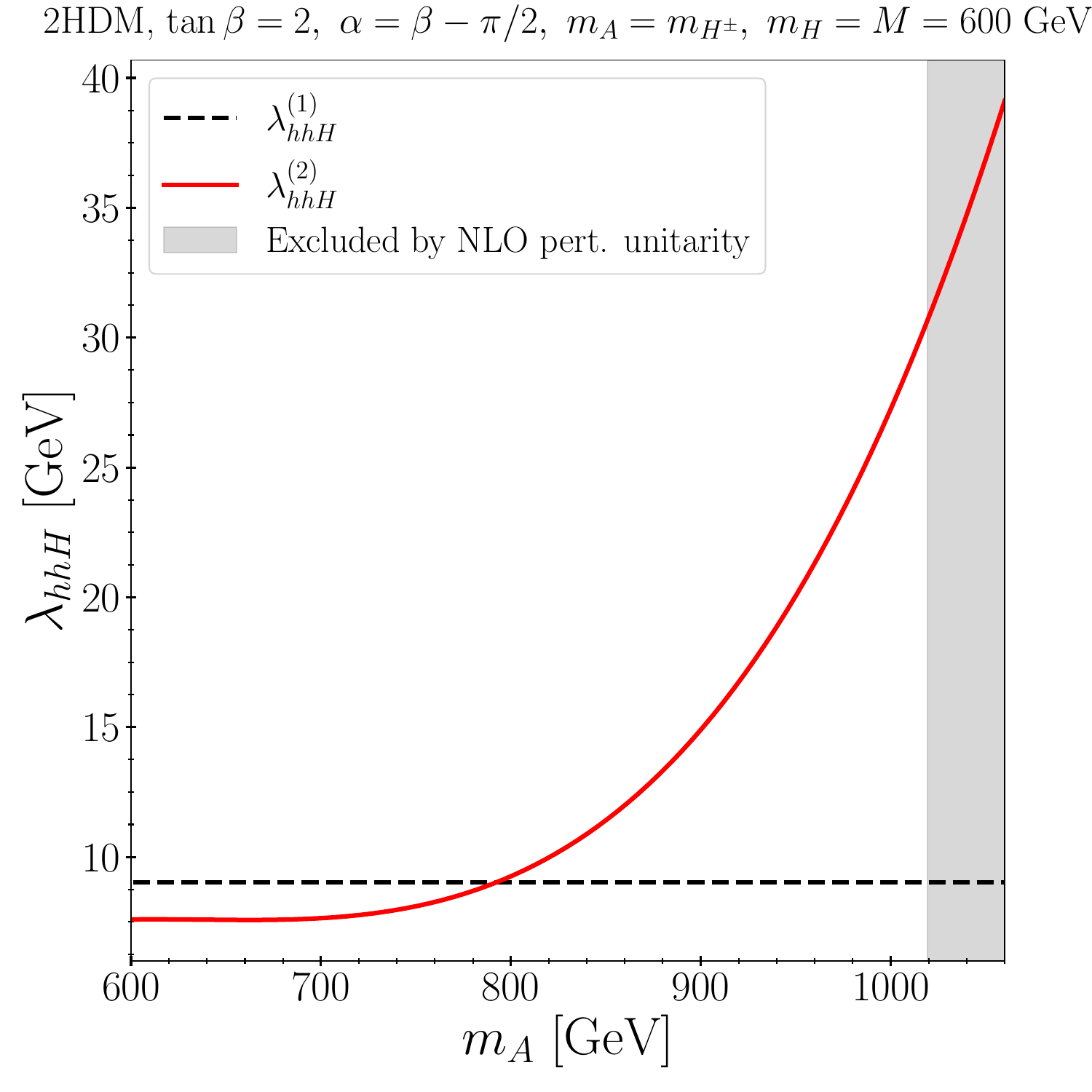}
    \caption{Results for $\kappa_\lambda$ (\textit{left}) and $\lambda_{hhH}$ (\textit{right}) in a 2HDM benchmark scenario with $M=m_H=600$ GeV, $\tan\beta=2$, $\alpha=\beta-\pi/2$ and varying $m_A=m_{H^\pm}$. Black lines indicate results at the leading one-loop level, while the red solid lines are our new results including leading one- and two-loop corrections to the trilinear scalar couplings. The blue solid line in the left plot for $\kappa_\lambda$ correspond to the result of Refs.~\cite{Braathen:2019pxr,Braathen:2019zoh} where loop-induced deviations from the alignment condition were neglected. }
    \label{fig:BP1}
\end{figure}

\begin{figure}[ht]
    \centering
    \includegraphics[width=0.49\textwidth]{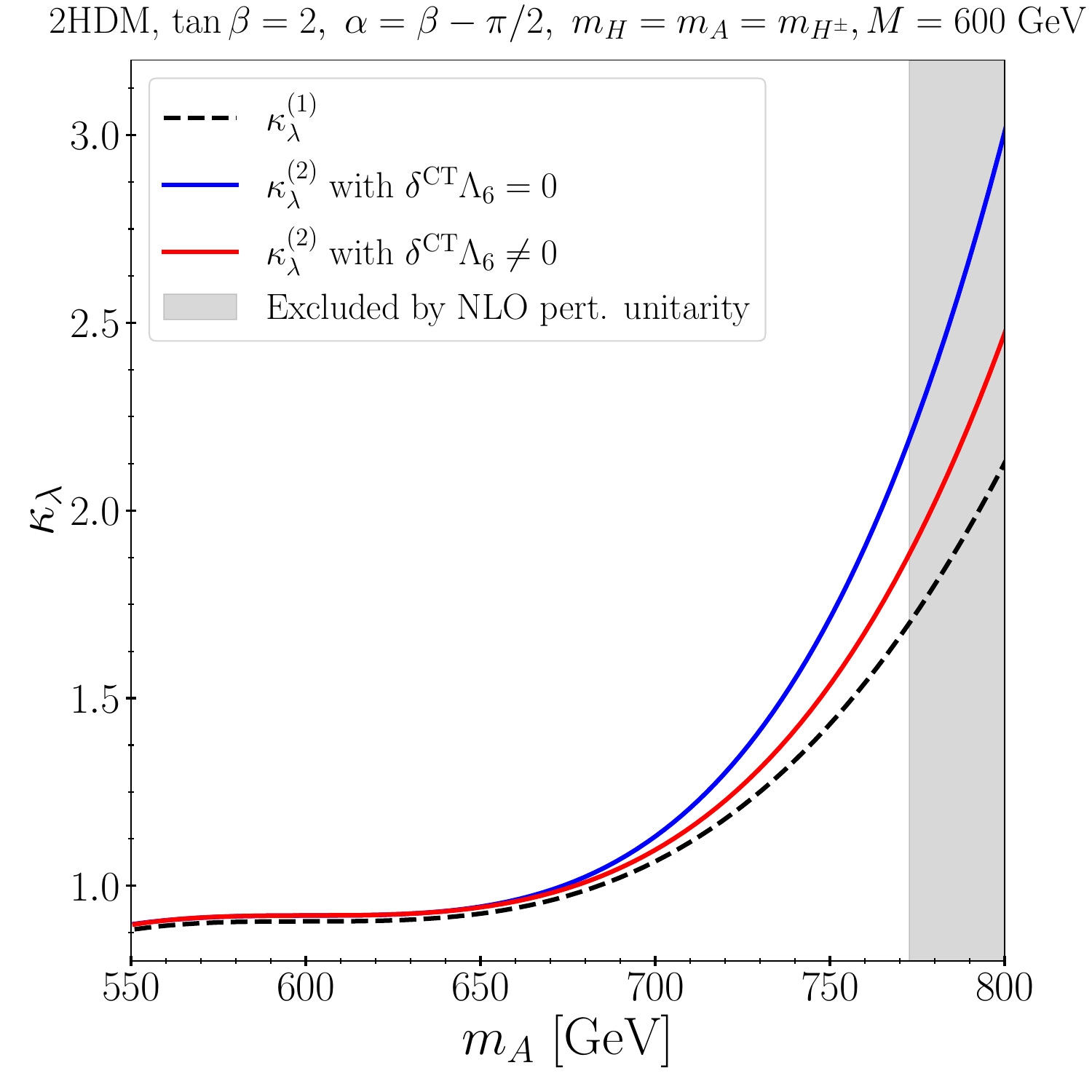}
    \includegraphics[width=0.49\textwidth]{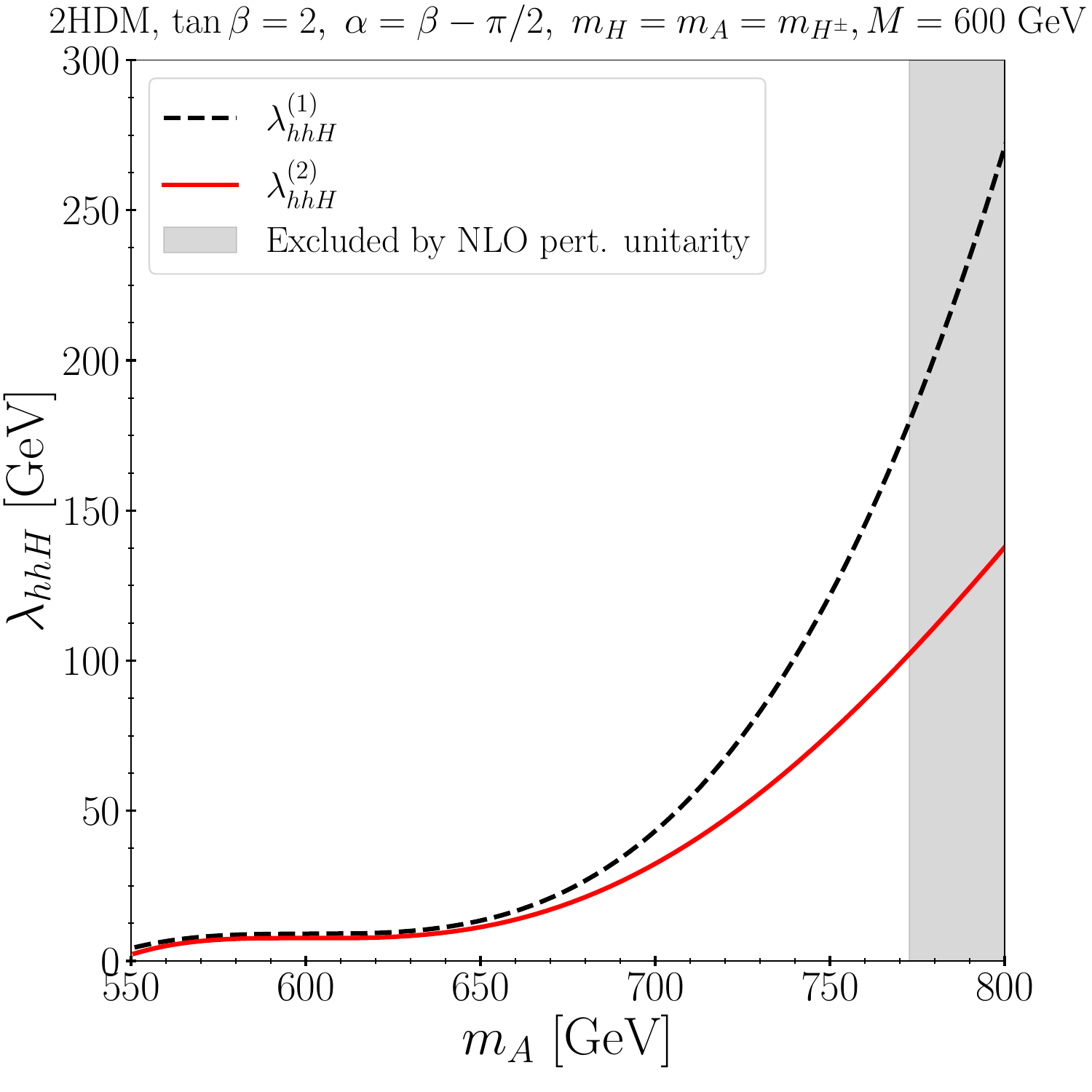}
    \caption{Results for $\kappa_\lambda$ (\textit{left}) and $\lambda_{hhH}$ (\textit{right}) in a 2HDM benchmark scenario with $M=600$ GeV, $\tan\beta=2$, $\alpha=\beta-\pi/2$ and varying $m_H=m_A=m_{H^\pm}$. The colour coding of curves is as in \cref{fig:BP1}.}
    \label{fig:BP2}
\end{figure}

In this section, we turn to numerical examples of our two-loop results for $\lambda_{hhh}$. 
We define the considered benchmark scenarios in terms of inputs 
in the $\mathbb{Z}_2$-symmetric basis, and we use the relations
\begin{align}
M_{22}^2=M^2 \,, \quad \Lambda_2=\frac{4(m_H^2-M^2)\cot^22\beta}{v^2}\,,\quad \Lambda_7 = \frac{2(m_H^2-M^2)\cot 2\beta}{v^2} \,, 
\end{align}
to relate these to parameters in the Higgs basis. We furthermore note that these relations are here written out in the alignment limit and with $m_h\to0$. 
Specifically, we consider two different scenarios: in \cref{fig:BP1}, we fix $M=m_H=600\text{ GeV}$ and vary $m_A=m_{H^\pm}$, while in \cref{fig:BP2}, we fix $M=600\text{ GeV}$ and vary $m_H=m_A=m_{H^\pm}$. For both scenarios, we choose $\tan\beta=2$ as well as the tree-level alignment condition $\alpha=\beta-\pi/2$. The first case is directly taken from fig.~2 of Ref.~\cite{Bahl:2022jnx}, whereas the second is inspired by the numerical investigations in Ref.~\cite{Braathen:2019zoh}. The left (right) panels of \cref{fig:BP1,fig:BP2} display the results for $\kappa_\lambda$ ($\lambda_{hhH}$); the black dashed line represents the (leading) one-loop prediction, while the red solid line indicates our new results including also leading two-loop corrections with the full OS renormalisation scheme discussed in the previous section. In the plots for $\kappa_\lambda$, the blue solid line corresponds to the result employing the expressions from Refs.~\cite{Braathen:2019pxr,Braathen:2019zoh}, where the renormalisation of $\alpha$ (or equivalently $\Lambda_6$) is formally performed in the \MS scheme --- this implies that the interpretation of $\alpha$ differs between the red and blue curves and a scheme conversion (in the one-loop contributions) would in principle be needed to compare the two results. The grey shaded regions are excluded by the theoretical constraint of perturbative unitarity, for which we use the next-to-leading-order (NLO) results from Refs.~\cite{Cacchio:2016qyh,Grinstein:2015rtl}. 

In the first scenario, in \cref{fig:BP1}, we can observe that the results with and without the finite part of the $\Lambda_6$ counterterm are almost indistinguishable. The small impact of the $\Lambda_6$ counterterm can be explained by the fact the dominant contribution therein is proportional to $(M^2-m_H^2)$, and thus vanishes in this particular scenario. This demonstrates that the approximation of neglecting deviations from the alignment condition at tree level is well justified in this case, and the conclusions of Ref.~\cite{Bahl:2022jnx} are unchanged --- both qualitatively and quantitatively. The two-loop results for $\lambda_{hhH}$ are also particularly interesting: indeed, setting $M=m_H$ implies that all the couplings $\lambda_{hHH}$, $\lambda_{HHH}$, $\lambda_{HAA}$ and $\lambda_{HH^\pm H^\mp}$ vanish at tree level, and in turn, only the top quark loop contributes to $\lambda_{hhH}$ at one loop (while we recall that $\lambda_{hhH}$ also vanishes at tree level due to the alignment condition). Therefore, radiative corrections to $\lambda_{hhH}$ involving the BSM scalars only appear from the two-loop level. For large mass splittings, $m_A-m_H\gtrsim 300\text{ GeV}$, the two-loop corrections can cause a significant deviation from the one-loop result. We emphasise that this does not by itself indicate a breakdown of perturbative unitarity as these two-loop effects involve couplings not present at one loop and are thus not a direct perturbative correction of the one-loop result, but instead a new class of contributions appearing from two loops and onward. 

Turning next to the second scenario, we find in the left plot of \cref{fig:BP2} that the effect of the finite $\Lambda_6$ counterterm is much more significant in this case, as could be expected given that $M\neq m_H$. The proper OS treatment of $\Lambda_6$ (or equivalently $\alpha$) leads to a reduction of the magnitude of the two-loop corrections to $\kappa_\lambda$. For $\lambda_{hhH}$, the one-loop contributions are significantly larger than in the previous scenario (given that one-loop diagrams involving BSM scalar do not vanish here).
Moreover, the two-loop corrections to $\lambda_{hhH}$ reduce the magnitude of the loop corrections, compared to the one-loop result. This reduction is stronger for
increasing scalar masses and therefore larger couplings. Within the allowed parameter region, before perturbative unitarity breaks down, the absolute size of the two-loop corrections are smaller than the one-loop correction, indicating perturbative convergence.

\section{Phenomenological impact on di-Higgs production}

In this section, we consider the calculation of physical observables involving the trilinear scalar couplings $\lambda_{hhh}$ and $\lambda_{hhH}$, specifically the total cross-sections and differential $m_{hh}$ distributions for Higgs pair production in the process $gg\to hh$ at the (HL-)LHC. We compute predictions for these at leading order in QCD (including however a $K$-factor of 2 to approximate the NLO QCD corrections), but with the dominant BSM corrections to the trilinear scalar couplings computed up to the two-loop level, using the public code \texttt{anyHH}~\cite{Bahl:2026anyhh,Bahl:2026nsu}.

\begin{figure}
    \centering
    \includegraphics[width=0.49\textwidth]{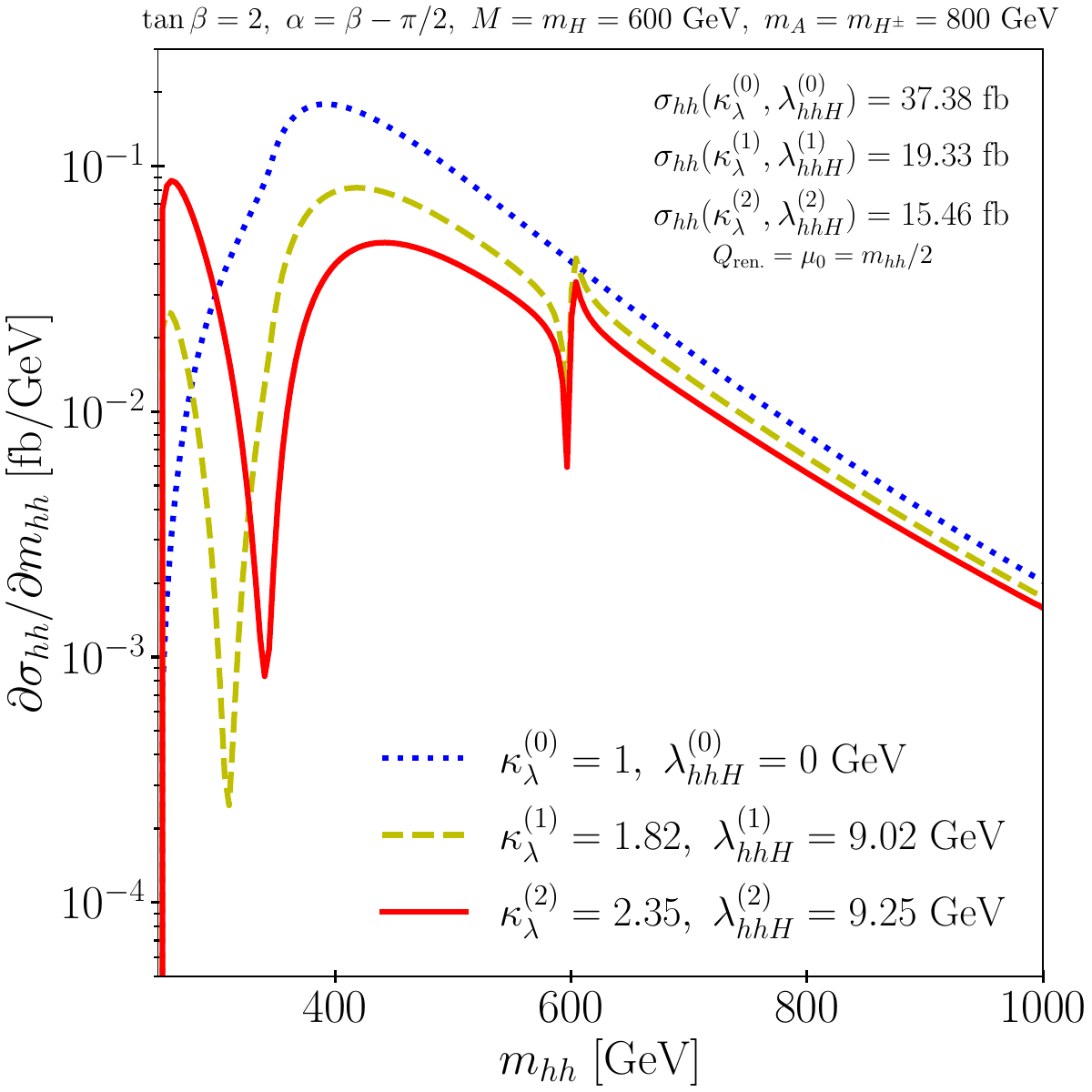}
    \includegraphics[width=0.49\textwidth]{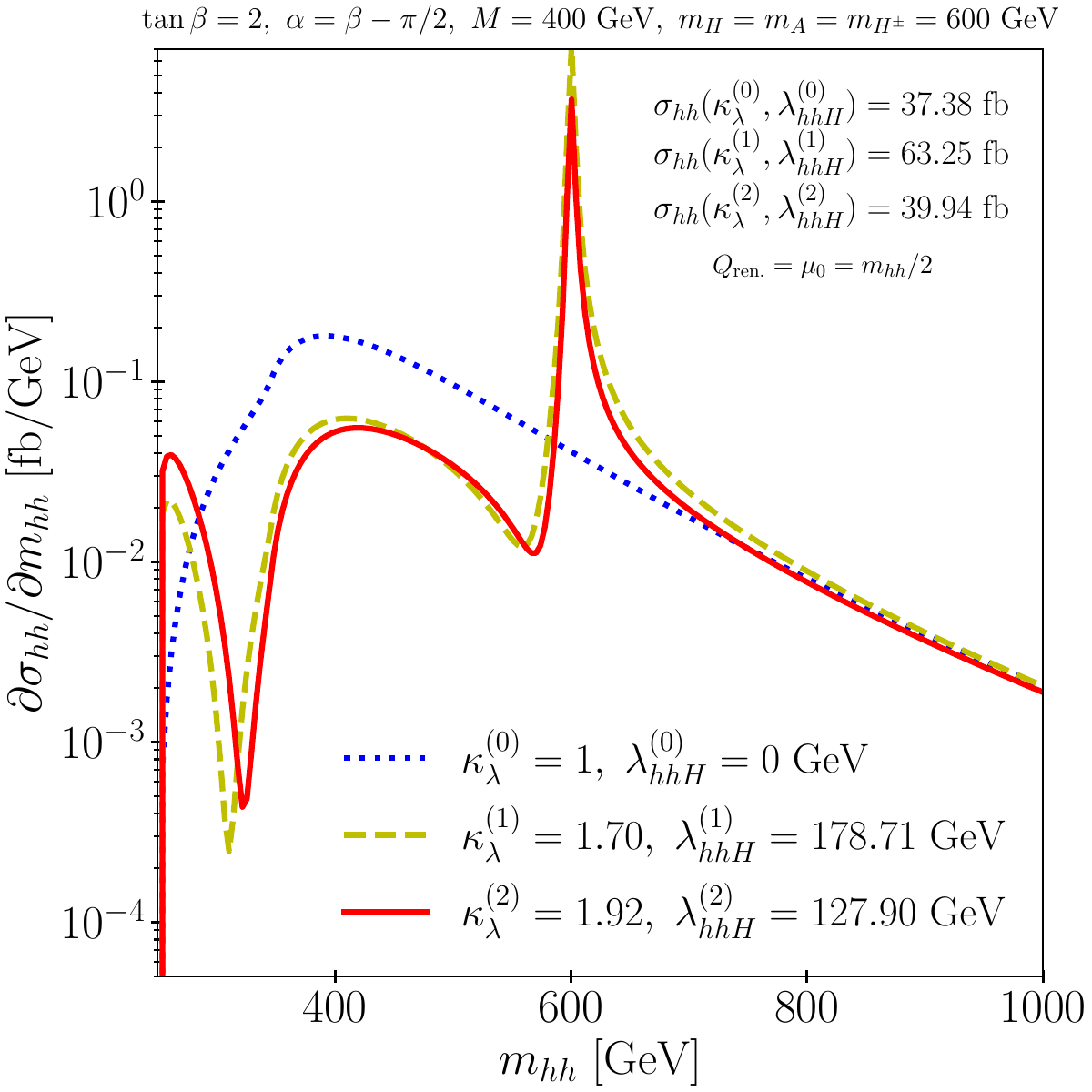}
    \caption{Differential cross-sections for Higgs pair production at the (HL-)LHC as a function of the di-Higgs invariant mass $m_{hh}$. A QCD $K$-factor of 2 is used to approximate the contribution of NLO QCD corrections, and the PDF factorisation scale is set to $\mu_0=m_{hh}/2$. \textit{Left:} results for BP1 defined in \cref{eq:bp1}; \textit{right:} results for BP2 defined in \cref{eq:bp2}. The different lines correspond to the distributions calculated with trilinear scalar couplings computed at the tree level (dotted blue), one-loop (dashed yellow) and two-loop (solid red) orders.}
    \label{fig:hh}
\end{figure}

We choose two benchmark points taken from the two scenarios in the previous section, and featuring resonances at $m_H=600\ \text{GeV}$. Specifically, we define
\begin{align}
\label{eq:bp1}
\text{BP1: }\quad    M=m_H=600\text{ GeV},\ m_A=m_{H^\pm}=800\text{ GeV},\ \tan\beta=2,\ \alpha=\beta-\frac{\pi}{2}\,,
\end{align}
and
\begin{align}
\label{eq:bp2}
\text{BP2:}\quad    M=400\text{ GeV},\ m_H=m_A=m_{H^\pm}=600\text{ GeV},\ \tan\beta=2,\ \alpha=\beta-\frac{\pi}{2}\,.
\end{align}
The differential cross-sections for these two BPs are shown in \cref{fig:hh} as a function of the di-Higgs invariant mass $m_{hh}$. The blue dotted, yellow dashed and red solid curves correspond to the predictions obtained with the trilinear scalar couplings computed at tree level, one loop and two loops (respectively). Results for the total cross-sections, using trilinear scalar couplings at different orders, are provided in the upper right corners of both plots.

As observed already e.g.\ in Refs.~\cite{Heinemeyer:2024hxa,Braathen:2025qxf}, the inclusion of loop corrections to the trilinear scalar couplings (i.e.\ the change from the blue dotted to the yellow dashed lines) can have a drastic impact on the $m_{hh}$ distributions, also in scenarios with alignment like those considered here.\footnote{We note that in Refs.~\cite{Heinemeyer:2024hxa,Braathen:2025qxf}, the differential distributions were studied with full one-loop calculations of the trilinear scalar couplings, not making any simplifying approximations (like $m_h\to 0$ or the gaugeless limit). Large effects were also found for points away from the alignment limit, see e.g.\ Ref.~\cite{Braathen:2025qxf}.} The large BSM effects in $\kappa_\lambda$ significantly modify the interference pattern between the box and SM-like triangle diagrams near the di-Higgs production threshold at $m_{hh}\gtrsim 250\text{ GeV}$. Furthermore, the non-zero value of $\lambda_{hhH}$ generated at the loop level leads to the appearance of a resonant contribution (with a dip-peak structure) around $m_{hh}=m_H$. 

The inclusion of the dominant two-loop corrections to the trilinear couplings does not lead to a new qualitative change of the distributions, although it has a notable impact at the quantitative level. Most significantly, the two-loop corrections further increase the BSM deviation in $\kappa_\lambda$, which causes an extra shift of the location of the maximal destructive interference between box and triangle contributions towards higher values of $m_{hh}$ (i.e.\ away from the threshold). The impact of the two-loop corrections on the resonance is less pronounced, especially for BP1, although for BP2 the width and height of the peak are affected by the $\sim 30\%$ decrease in $\lambda_{hhH}$. 

The impact of the two-loop corrections to $\lambda_{hhh}$ and $\lambda_{hhH}$ on the total cross-sections is of the order of $-20\%$ for BP1 and of $-37\%$ for BP2. This can be understood from the combined decreases caused by the change of the interference pattern\footnote{We recall in this context that the minimum of the di-Higgs cross-section predictions with floating $\kappa_\lambda$ occurs around $\kappa_\lambda\sim 2-2.5$~\cite{LHCHiggsCrossSectionWorkingGroup:2016ypw}.} at the threshold --- which is most pronounced for BP1 --- and the effect on the resonant peak --- which is largest in BP2. Overall, these results again indicate the perturbative convergence of the calculations, but also still demonstrate the relevance of including higher-order corrections to trilinear scalar coupling in precision predictions for the Higgs pair production process.

\section{Conclusions}
In this work, we have presented updated results for the dominant two-loop corrections, involving BSM scalars and the top quark, to the trilinear scalar couplings $\lambda_{hhh}$ and $\lambda_{hhH}$. These improve on the early results of Refs.~\cite{Braathen:2019pxr,Braathen:2019zoh} by including a proper on-shell renormalisation of the mixing angles, and by extending calculations to BSM trilinear couplings (following similar improvements already obtained in Ref.~\cite{Degrassi:2025pqt}). Moreover, our results open the way to the first full on-shell renormalisation of general 2HDMs in the (Higgs) alignment limit beyond one loop. In particular, we have considered an on-shell renormalisation of the alignment condition up to the two-loop level, and discussed its impact on the two-loop calculation of $\lambda_{hhh}$ in comparison with earlier results, and considering two scenarios inspired by the recent literature. We have also investigated the impact of these corrections on total cross-sections and differential distributions for Higgs pair production at the (HL-)LHC computed with \texttt{anyHH}.  In this context, we have pointed out that while perturbative convergence appears to be ensured, the dominant two-loop BSM corrections to the trilinear scalar couplings still have a noticeable impact on di-Higgs predictions, thereby motivating their inclusion whenever possible. 

Further details and results from our calculations will be provided in future work. 

\subsection*{Acknowledgements}
We thank Kateryna Radchenko Serdula for many helpful discussions and exchanges about \texttt{anyHH}. We are supported by the DFG Emmy Noether Grant No.\ BR 6995/1-1.
We also acknowledge support by the Deutsche Forschungsgemeinschaft (DFG, German Research Foundation) under Germany's Excellence Strategy --- EXC 2121 ``Quantum Universe'' --- 390833306. This work has been partially funded by the Deutsche Forschungsgemeinschaft (DFG, German Research Foundation) --- 491245950.


\bibliography{bibliography.bib}

\end{document}